\documentclass[aps,prd,onecolumn,12pt,preprintnumbers]{revtex4-1}

\usepackage{hyperref}

\usepackage{amsfonts}

\usepackage{float}

\usepackage{tikz}

\usepackage{graphicx}
\usepackage{epstopdf}

\begin{document}



\preprint{HDP: 15 -- 05}

\title{\quad \quad \quad The Resonator Banjo Resonator, part 2:  \newline What makes 'em  ``really crack"?}

\author{David Politzer}

\email[]{politzer@caltech.edu}

\homepage[]{http://www.its.caltech.edu/~politzer}

\altaffiliation{\footnotesize 452-48 Caltech, Pasadena CA 91125}
\affiliation{California Institute of Technology}

\date{June 17, 2015}

\begin{figure}[h!]
\includegraphics[width=4.3in]{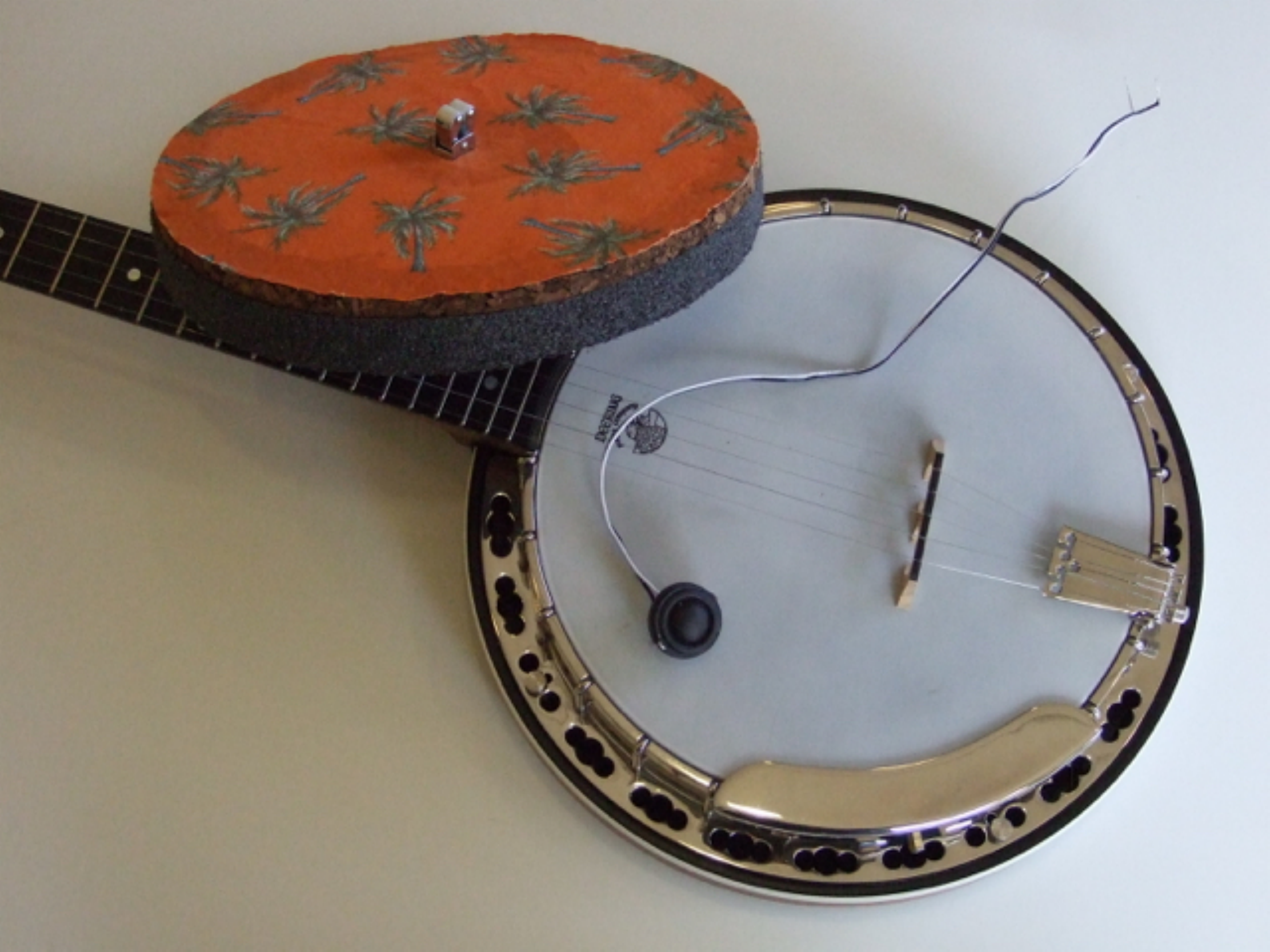}
\end{figure}

\begin{abstract}
A simple experiment quantifies the difference between the sound production of a banjo with and without a resonator back.  Driven by a small tweeter mounted inside the pot, for frequencies above about 4500 Hz, the produced external sound is 6 to 10 dB louder with the resonator than without.  With the banjo played in any normal fashion, this gives a negligible contribution to the overall volume.  However,  that difference is clearly a reflection of the universally recognized resonator sound, in close analogy to plosive consonants in human speech.  No direct correlation is observed between the head-resonator separation and the spectrum of the enhanced response.  This suggests that direct reflection off the back is not a primary contributor to the resonator/openback difference, leaving differences in overall absorption as the major suspect.   
\end{abstract}

\maketitle{ {\centerline{{ \bf \large The Resonator Banjo Resonator, part 2:   What makes 'em ``really crack"?}}}

\section{Introduction}
What is the role of the resonator on a resonator banjo?  A previous exploration\cite{part1} found that it doesn't make the banjo particularly louder overall --- as compared to the same instrument with the resonator removed.  This observation was well-known to some  players and regarded as obvious nonsense by others.  However, there is no disagreement that the presence of the resonator changes the sound, producing something with sharper projection, better articulation, and subtly different tone.  It allows the banjo to cut through the sound of other instruments.  Other aspects of the timbre change as well.  In its extreme version, it is said that a fine resonator banjo ``really cracks."

I couldn't identify any easily quantifiable, distinguishing features in sound recordings of normal playing --- not in the time sequence, its frequency spectrum, or any combination of the two.  The features common to resonator on {\it versus} off were too overwhelming.   But I thought of an experiment that might highlight the differences.

Normal banjo sound production proceeds as follows: The plucked string moves the bridge; the bridge moves the head; and most of the sound comes directly off the head.  The head also sets in motion the air inside the pot (the cylindrical volume between the head, rim, and back [resonator or player's belly]).  Some of the pot  sound comes out the sound hole (the space between the rim and back), and some reacts back on the head, altering its motion and, hence, the sound it produces.  Changing the back can change the way the pot air responds to being driven, which, in turn, effects the total produced sound.

\section{The Tweeter-driven Pot}
In ref.~\cite{part1}, simple measures of overall loudness did not distinguish a resonator from an openback played naturally or with an ersatz experimental-grade ``synthetic" belly (constructed of cotton fabric, cork, and closed-cell foam to approximate reflection and absorption of a real belly in a reproducible fashion).  The detailed analysis was restricted to below 4000 Hz.  In spite of the banjo being far richer in its high frequency production than other instruments, the power spectrum nevertheless drops dramatically with increasing frequency.  There's just not much power at higher frequencies.  However, in speech and song, frequencies much higher than musical fundamental pitches are crucial to understanding the words.  Consonants, particularly stops or plosives,
are very short-time features that don't appear in simple spectral analysis or energy accounting.  But our brains efficiently process their meaning.

I decided to look at external sound production driven by high frequency vibration introduced inside the pot.  I mounted a $3/4''$ tweeter inside a Deering Sierra, damped the strings, swept the tweeter slowly in frequency from 150 to 20,000 Hz at fixed amplitude.   I recorded with a microphone at $7''$ from the head, facing toward the plane of the head, but located above the edge.
More details of the set-up and discussion of the results are given later.
\begin{figure}[h!]
\includegraphics[width=4.0in]{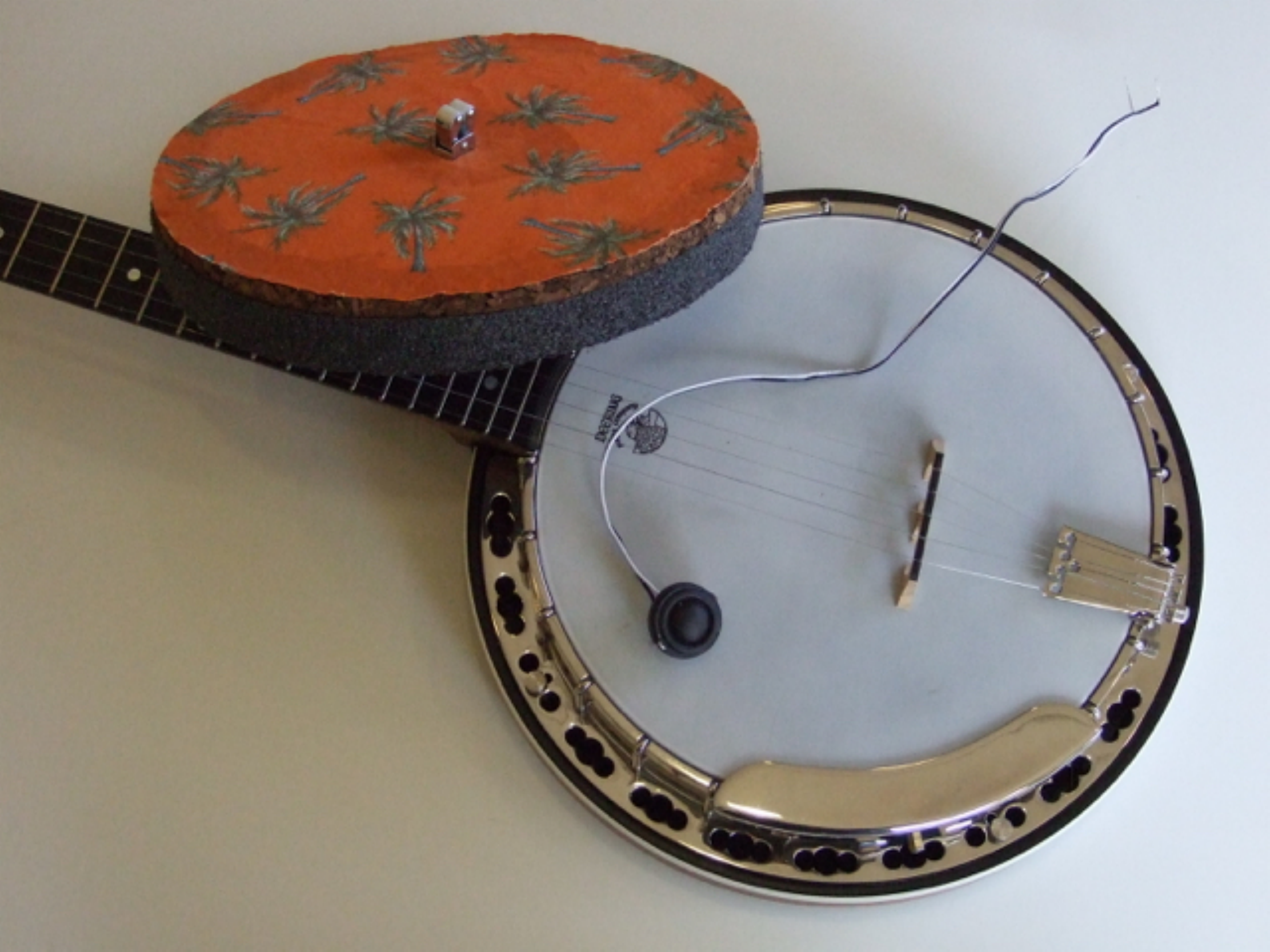}
\caption{Resonator banjo, $3/4''$ tweeter, and synthetic belly}
\end{figure}

The basic results are presented in FIG.~2.
For reference, the spectrum of the tweeter just by itself, as recorded at $7''$, is included in the graph.  The tweeter has trouble making much sound below 500 Hz.  Its response to a fixed amplitude voltage certainly isn't flat as a function of frequency, but it does roughly match the manufacturer's specs.  And it does a reasonable job over the range of interest.  The tweeter spectrum looks nothing like the spectrum of a plucked string at the bridge, the spectrum of normal bridge motion, or the spectrum of any portion of the head when the banjo is played.  These fall dramatically with increasing frequency.  But the tweeter drive allows an investigation of otherwise subtle high-frequency aspects of the pot sound production.
 
\begin{figure}[h!]
\includegraphics[width=4.95in]{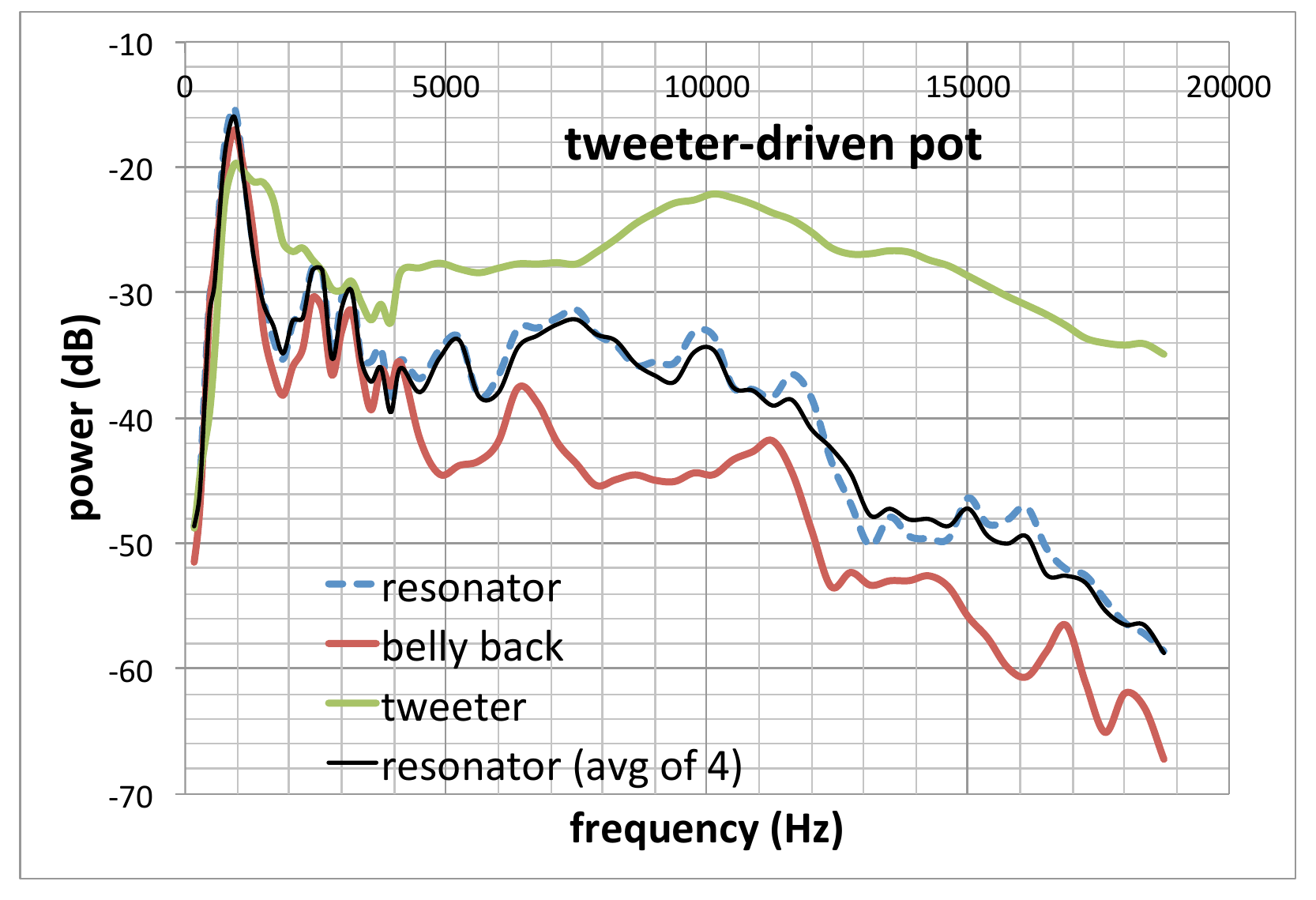}
\caption{sound production of the tweeter frequency sweep}
\end{figure}

\bigskip

\begin{figure}[h!]
\includegraphics[width=4.85in]{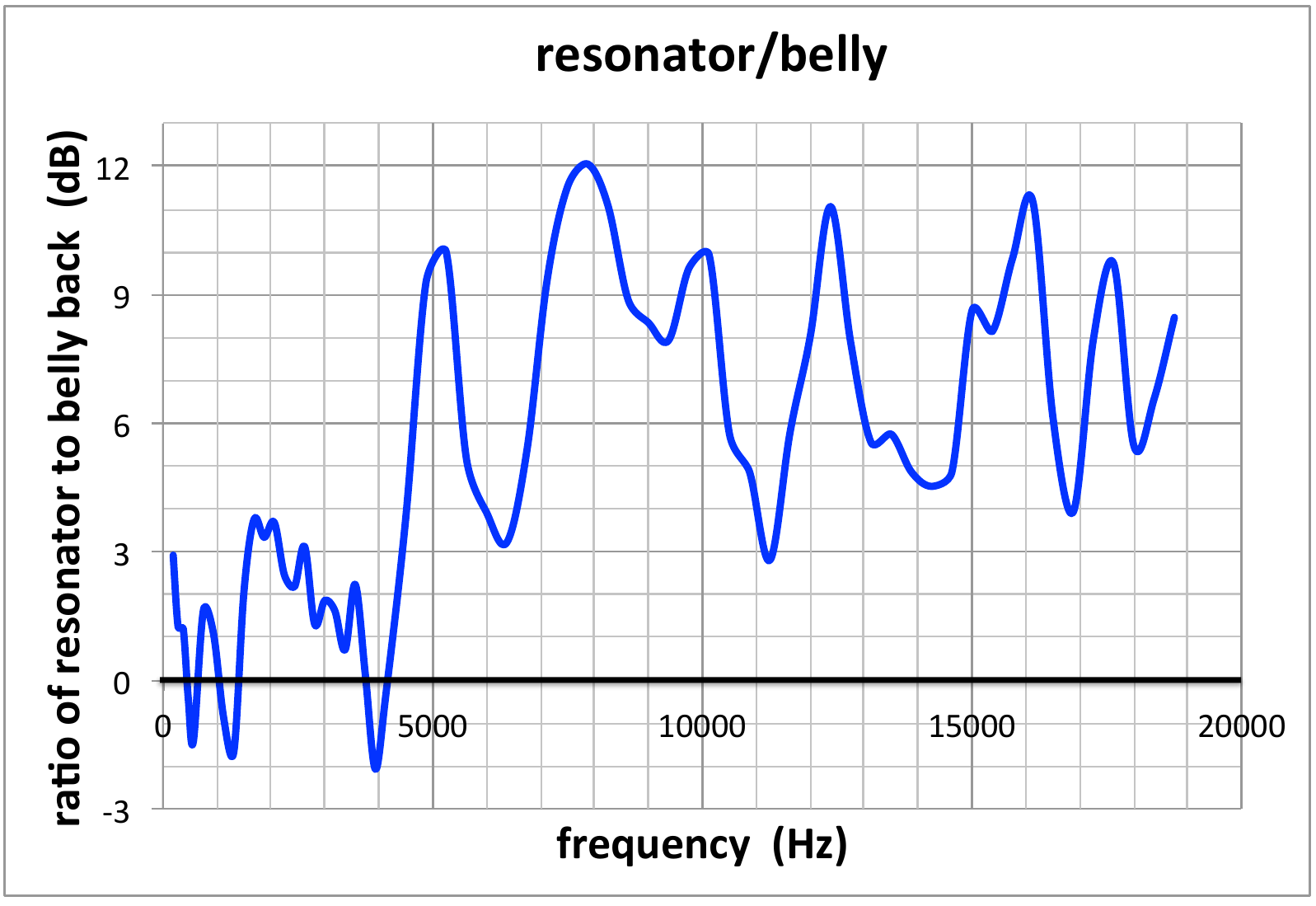}
\caption{resonator-belly comparison as a ratio of spectra}
\end{figure}

The dashed curve in Fig.~2 is a single sweep with the resonator back, and the solid black curve is the average for four such sweeps.  The latter is naturally the smoother of the two, but they rarely differ by more than 2 dB.  (More on this later.)  The red (lighter) curve is a sweep with the synthetic belly back.  (See ref.s~\cite{part1} and \cite{openback} for belly details.)  Between 1400 and 3700 Hz, the resonator back is typically about 2 dB louder.  That kind of difference is often considered barely perceptible.  However, starting at about 4400 Hz, the difference between the two backs is dramatic and remains so over the entire high frequency range.

To emphasize the identification of the difference between the resonator and belly back, I plot their ratio in FIG.~3.  Decibels give a logarithmic measure, and the plot is of the decibel differences between backs.  In terms of the actual microphone signal as an oscillating voltage, what is calculated is the ratio of the spectra --- and then that ratio is expressed in decibels.  A zero decibel difference indicates equal strength signals.

\section{the lesson}
Apparently, the resonator improves the produced sound response to high frequency excitation of the pot.  But we do not listen to musical pitches in that high frequency range.  In fact, it is difficult (or even  impossible) to distinguish the pitch of pure sinusoidal tones above some frequency.  For me, that's anything over about 5000 Hz.  That frequency and above are quite audible, but they just sound like identical squeals.  At some much higher frequency, pure tones become totally inaudible.  However, these frequencies contribute to the sound and timbre of musical notes.  Likely, even much of the ``inaudible" range contributes to distinctions in sound that we can hear.  Perception of sound is not simply a matter of frequency decomposition.  
Rather, we simultaneously process our auditory inputs in a variety of ways.  We consider both the time sequence and frequency content.  We evaluate intervals of various lengths in each (rather than simply point-by-point functions\cite{prl-non-lin}).  And these parallel lines of analysis exchange partial results along the way --- enhancing the effectiveness of each.  We are certainly aware of and can distinguish features of sound that rely on very high frequency components, such as dramatic changes that occur very quickly.

One can divide the pot interior's impact on sound into two parts: 1) its reaction back onto the head and 2) the sound radiated from the sound hole.  The interaction between the head and the air inside the pot is quite complex.  They are in direct contact, pushing on each other, over a dynamic surface.  The very lowest and also the asymptotically highest frequencies are simple to picture, but it is the intervening ones that are relevant to music.  Nevertheless, it is clear that the resonator back dissipates {\it considerably} less vibrational energy above 5000 Hz than a belly.  So more energy is available for sound.  In the pure time domain description, the resonator back enables a more faithful produced-sound response to the driving by the plucked strings.

Elementary acoustical science tells us about the sound radiation from sound holes in two limiting cases.  If the sound wavelength is much longer than the sound hole's dimensions, then sound is radiated isotropically but only {\it very} weakly.   If the sound wavelength is much shorter than the sound hole's dimensions, then sound is narrowly beamed in the forward direction and is of relatively strong amplitude.  Inbetween is inbetween, but that's where banjos live.  100 to 20,000 Hz corresponds to wavelengths between 10 feet and 11/16 of an inch.  Its high frequency strength gives the resonator banjo an advantage over the open back both in terms of directionality of projection and the amount of direct radiation from the sound hole.  The directionality aspect is probably why resonator banjo design has settled for the past century on the forward facing sound hole rather than sideways, as provided by a simpler flat disk back spaced slightly off the bottom of the rim.

\section{comparisons of equalized recordings}
Doing a resonator banjo justice requires some decent picking.  Ken LeVan, designer and luthier extraodinaire\cite{levan}, agreed to let me use some of his sound files.  In particular, a short selection of {\it Pretty Polly} is presented here, with and without a resonator on the same banjo, recorded as identically as possible.  Of course they sound different --- although the differences might only be immediately apparent to people who listen a lot to banjos.

To test the relevance of the tweeter drive results to actual music listening, I constructed an ``equalizer" in Audacity{\small $^{\circledR}$} that is a cartoon version of the resonator enhancement shown in FIG.~3.  That equalizer can be applied to any sound file to create a new one that is enhanced, as a function of frequency, by the specified amount.  So I created  an enhanced version of Ken's openback rendition.  The listener can decide whether this gives the performance the extra punch and sparkle characteristic of a resonator banjo.  Conversely, I wrote an equalizer function that is the inverse of the FIG.~3 resonator/belly difference and applied it to Ken's resonator back performance.

That makes four versions of the same short tune.  The files are 
\href{http://www.its.caltech.edu/~politzer/ppA.mp3}{ppA.mp3},  \href{http://www.its.caltech.edu/~politzer/ppB.mp3}{ppB.mp3}, \href{http://www.its.caltech.edu/~politzer/ppC.mp3}{ppC.mp3}, and \href{http://www.its.caltech.edu/~politzer/ppD.mp3}{ppD.mp3}.  (Click on those file names or retrieve them from \href{http://www.its.caltech.edu/~politzer/}{Banjo Physics 411}.)  Can you hear the differences and identify which is which?  Good speakers help.  The answer is at the end of this paper, just before the notes and references, in Appendix B.  Note that the equalizer values came from a comparison of a Deering Sierra with and without its resonator, and it was a crude approximation at that.  The sound files are played on a LeVan one-of-a-kind.  Further adjustment of the equalizer form might have been able to bring the corresponding real and equalized recordings even closer.  The question for now is whether the enhanced openback plausibly sounds like a resonator banjo and the suppressed resonator sounds like an openback.

\section{the equalized sound, plosives, and voice recognition}
The application of the equalizers defined by FIG.~3 to real, recorded music files highlights a conundrum.  An openback is made to sound plausibly like a resonator banjo and {\it vice versa}.  However, FIG.~3 represents the sound produced by air oscillations in the pot.  The observed many decibel difference at high frequency is a feature of the pot and its backs, not of the whole banjo sound.  Most of banjo sound comes directly off the head.  A spectrum analysis of the original, complete, unaltered music recordings reveals only slight differences between the backs, in spite of their being identifiably different to a listener.  Processing one of the recordings through the equalizer produces something that sounds much closer to the other recording.  But now the difference in spectra is no longer slight.  It is roughly given by FIG.~3.

A plausible resolution of the disparity is the following.  Those high frequencies play at least two idistinguishable roles.  1) They contribute to high frequency ripples superposed onto lower frequency oscillations, and 2) they combine with each other to produce features in the sound that are sharp as a function of time.  For very high frequencies, we take notice of 2) but are barely aware of 1).  The equalizer enhances the strength of 2) in the openback recording.  It also enhances the strength of 1), but that goes largely unnoticed --- and conversely, going the other way.

An enlightening simple exercise with spoken words can be performed with a modern computer and sound software.  Record the eight, three letter words formed by \{$T$ or $D$\}, \{$I$ or $A$\}, and \{$P$ or $B$\}.  Multiple versions of each would help.  Those four consonants are plosives, characterized by high pressure pulses of very short duration.  Played back, just by themselves, they can be hard to distinguish.  Furthermore, a Fourier decomposition of just those short durations has significant power between 6000 and 20,000 Hz.  Pure sinusoidal tones of the lower end of that range of frequencies are indistinguishable to most people's ears, and the higher end is simply inaudible.  Nevertheless, those frequencies are essential to the understanding of speech.

Computerized voice recognition plays a variety of roles in current society, some sinister and some not.  The recognition of vowel sounds goes back to Helmholtz.  He had his wife sing vowel sounds into a piano, and he carefully noted which strings resonated.  He identified the characteristic patterns of string excitation that we now call ``formants."  Fricatives are the consonants that have extended sounds.  Examples are $f$, $j$, $l$, $m$, $n$, $s$, $sh$, $v$, and $z$.  They come with their own enhanced regions of frequency, all rather higher than the vowels.  Plosives are the short ones, nearly instantaneous in time.  They don't involve much sound energy, relative to the others.  We hear them nevertheless.

Plosives were a challenge to computer recognition, but huge advances in computer power have allowed a practical solution.  A computer can identify all the possible plosives that might correspond to a particular sound.  The adjoining sounds will combine to make possible words.  If more than one is actually a word in the language in use, the computer can look at the containing phrase to narrow down which makes sense.  This might seem like an elaborate algorithm.  But computers do billions of steps per second.  Our brains might be doing something very similar.  While we only do steps at a rate of thousands per second, we can do zillions of such chains simultaneously.  (So we're still a bit ahead.)

\section{some simple physics}
Qualitative aspects of sound reflection and absorption by common materials are part of everyday experience.  Elementary physics has little to add.  Only the most general features and trends can be explicated.  Orderly crystals can be hard to distort, and, when they do, they do so ``elastically," giving back the energy that had been put into the distortion.  Tangles of very long molecules might typically distort more easily and lose energy to the friction between them.  The details are consequences of molecular dynamics and are the subject of practical studies in acoustical engineering.

But it would be helpful to understand a bit about what the sound is ``doing" inside the pot.  ``Ballistic" or ``ray" trajectory descriptions, familiar from thinking about light and common objects, can be misleading (unless one takes care to keep track of phase information along the paths\cite{feynman}).  They really only make sense if the wavelength is much shorter than the object's dimensions.  And, even then, edges of objects are inherently very short scale features, irrespective of the overall size.  So wave-like features can appear due to edges of large objects.  For wavelengths much longer than any of the object's dimensions, the shape of the object becomes irrelevant.

A typical physics approach is to consider the ``normal modes," i.e., resonances and resonant motions, of the air inside the pot.  This is useful if that air motion is only weakly coupled to other parts of the system, including both coupling to other oscillations and to dissipative damping effects.  In such cases, the actual air motion is well-approximated by a superposition of the normal mode motions, at least for a time lasting several cycles of the relevant modes.  The case at hand, as with many cases in musical acoustics, seems to satisfy this approximation.\cite{tone-ring}  Coupling to the head and damping of the pot air motion are important, but they can be considered separately, after the basic air motion is understood.

The sound modes of a cylindrical cavity are well known.  They are labeled by three integers that count the number of nodal planes\cite{nodes} of three basic types.  There are concentric cylinders, diameter planes equally spaced in angle, and flat disks parallel to the cylinder top and bottom.  A given mode has some number of each.  If the cavity is long and thin (like the pipe of a wind instrument), the umpteen lowest frequency modes have no cylinder or diameter nodes.  All the nodes are equally spaced disks along the length.  For the squat banjo pot, it's the other way around.  The lowest frequency modes have cylinder and diameter plane nodes.  The lowest frequency mode with a disk node plane has a wavelength in free air given by twice the pot depth.   Half of a wavelength fits in the pot from top to bottom.  This is a standing wave, but it can be equally well described as the superposition of waves traveling back and forth between the head and the back, uniform in the plane perpendicular to that motion.  For a typical banjo, this motion is around 2500 Hz.  In the vicinity of that frequency, there are already many closely spaced resonances with radial and azimuthal variation (i.e., cylindrical and diametrical node planes).  Furthermore, above 2500 Hz, that one disk mode gets ``dressed" with air motions in the other directions, to yield yet higher resonant frequencies.  A second disk nodal plane appears around 3750 Hz, and it, too, gets dressed with the other kinds of nodal planes.

This is the physicist's answer to when air motion involving bouncing off the back is of dynamical importance.  The answer is that the frequency must correspond to travel at the speed of sound of a wave back and forth between the head and the back.  And that gives a relatively high frequency that is inversely proportional to the pot depth.  At lower frequencies, the pot air interaction with the back is also significant, but the air behaves as if it were not compressible in that direction.  Lower frequency head motions  force air in and out the sound hole and/or push it radially and/or from side-to-side.  

As shown in FIG.~3, 4400 Hz was where the resonator and belly backs began to differ appreciably.  That corresponds to a wavelength of 3 inches.  This is somewhat smaller than the back diameter.  Hence, it is possible that the shape (e.g., curvature) of the back has some relevance to the sound at that frequency and higher.  Plausible details are hard to picture, but the coupling of pot air motion to head motion might be affected by back curvature, particularly at the highest frequencies.

The shape of the resonator back can certainly have an very important impact for structural reasons, independent of the directionality of reflection of high frequency sound.  The resonator back wood is typically about $1/4''$ thick.  For the same wood and construction technique, a flat back will be more flexible than one with dish-shape curvature or ribs.  Flexing leads to greater dissipation.  So a curved back is a further contributor to the resonator goal.

To the extent that the geometries are the same or similar, the resonator-belly back difference is all about dissipation.  The normal mode description suggests a division of the general phenomenon into two parts: 1) air motion parallel to the back surface and 2) air motion perpendicular to the back surface.

1) Even well below 2500 Hz (or wherever the first disk node plane appears), the air motion parallel to the plane of the back will certainly lead to some energy dissipation in the back.  The important point is that the air motion also comes with a wave of pressure.  As a force, pressure is isotropic, i.e., the same on any plane that goes through a given point, irrespective of its orientation.  The pressure oscillations of the low frequency modes push and pull on the back, even if the basic air motion is parallel to the surface of the back.  This generates motion of the back material and consequent dissipation of energy.  In an idealized version of this situation with an interface between two infinite half-volumes, one of air and one of back material, waves in the air at vanishingly small angles to the interface do not generate traveling waves into the back material.\cite{total-internal}  However, there is motion of the back material that decays exponentially with distance from the interface.  That motion is subject to dissipation.

2) Air motion perpendicular to the back (or at least at greater than some critical angle from the surface) generates wave motion into the back material.  Energy loss to the air comes from the energy transferred  to motion in the back.  Some of that is dissipated nearby in the back.  If the back were, indeed, very thick, some of it would continue to travel to further depths, dissipating along the way. 

In conclusion, in addition to material property effects, there are geometrical effects.  For typical belly materials, sound reflection is weaker and absorption stronger at an air interface than for resonator materials.  And those differences increase with increasing frequency.  (These are observed acoustical engineering facts.)  But there is also a geometrical effect.  Dissipation by the back may become more significant when there is air motion perpendicular to the back.  In the language of Fourier analysis, that only happens above the frequencies of modes which include such motion.

\section{comparing different pot depths}
Through the generosity of Greg Deering, I happen to have three banjos that are as identical as possible as CNC milling can produce out of maple, differing only in the depth of their pots.\cite{openback}  In particular, they are research-grade Goodtime banjos with pot depths of $2''$, $2 {3 \over 4}''$, and $5 {5\over 8}''$.
\begin{figure}[h!]
\includegraphics[width=3.0in]{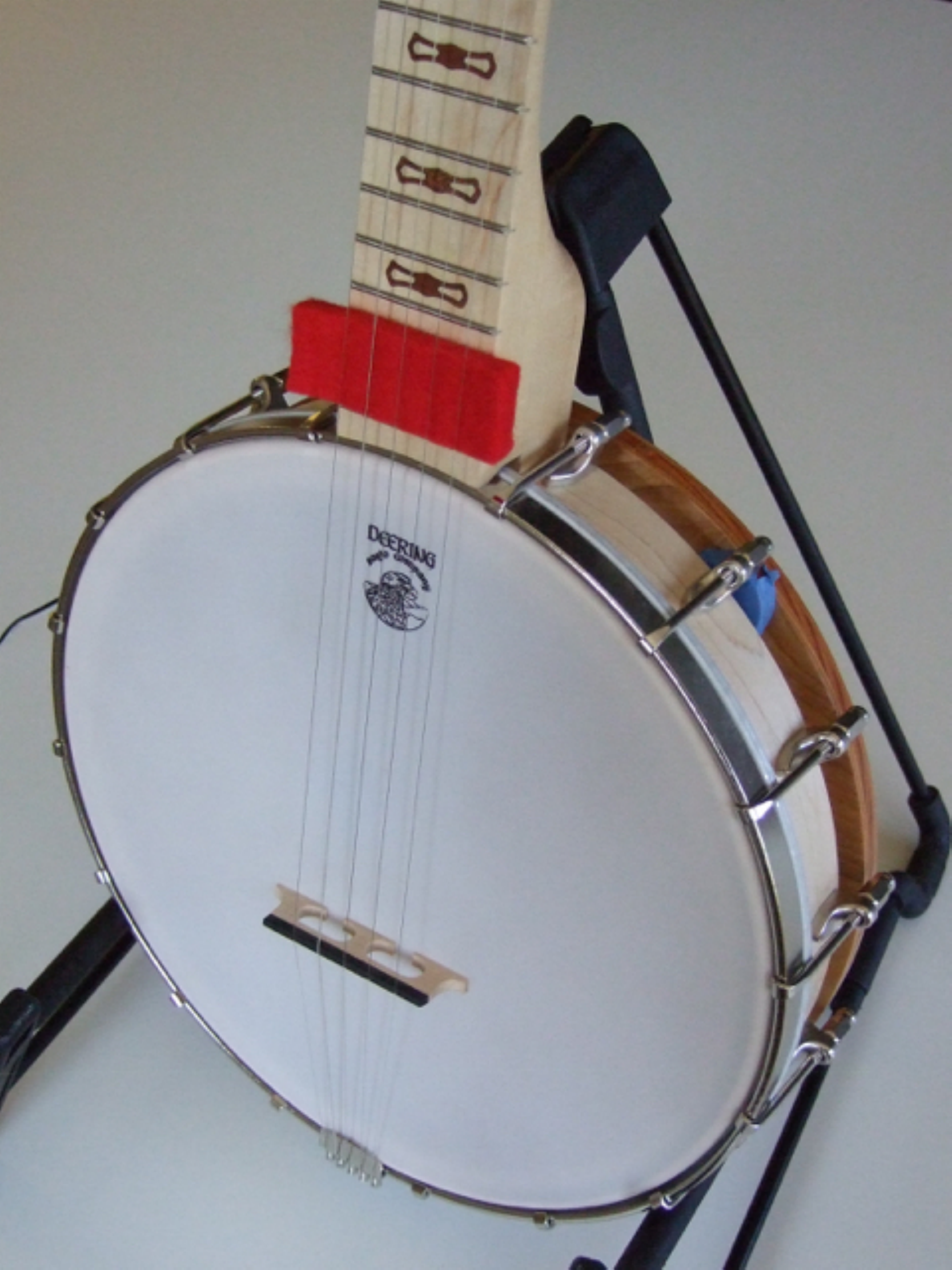}
\caption{$2''$ deep Goodtime with a wood disk back, spaced to match typical open-back playing\cite{openback}}
\end{figure}
This banjo trio allowed a study of how the distance between the head and the back impacts the difference between a resonator and a belly.  Mounting a traditional resonator on the three in an identical way is somewhat problematical.  Instead, I chose a simple alternative, which actually yields a better comparison.  In particular, I used the more primitive flat disk back, made of $1/2''$ plywood, mounted with the same clip-on hardware (cupboard latch to the coordinator rod) as the synthetic belly.  Hence, the geometry of the sound hole was identical for all banjos and backs.  This is in contrast to the comparison of the real resonator with the belly, where the difference in projection or radiation pattern was a confounding issue.  For these three banjos, each with two possible backs, the microphone was placed at $7''$ facing directly into the sound hole, i.e., facing the single $3/8''$ spacer that defined the angled opening sound hole of all of the backs.  The tweeter and the signal generater sweep were as they were for the Deering Sierra.

\begin{figure}[h!]
\includegraphics[width=6.5in]{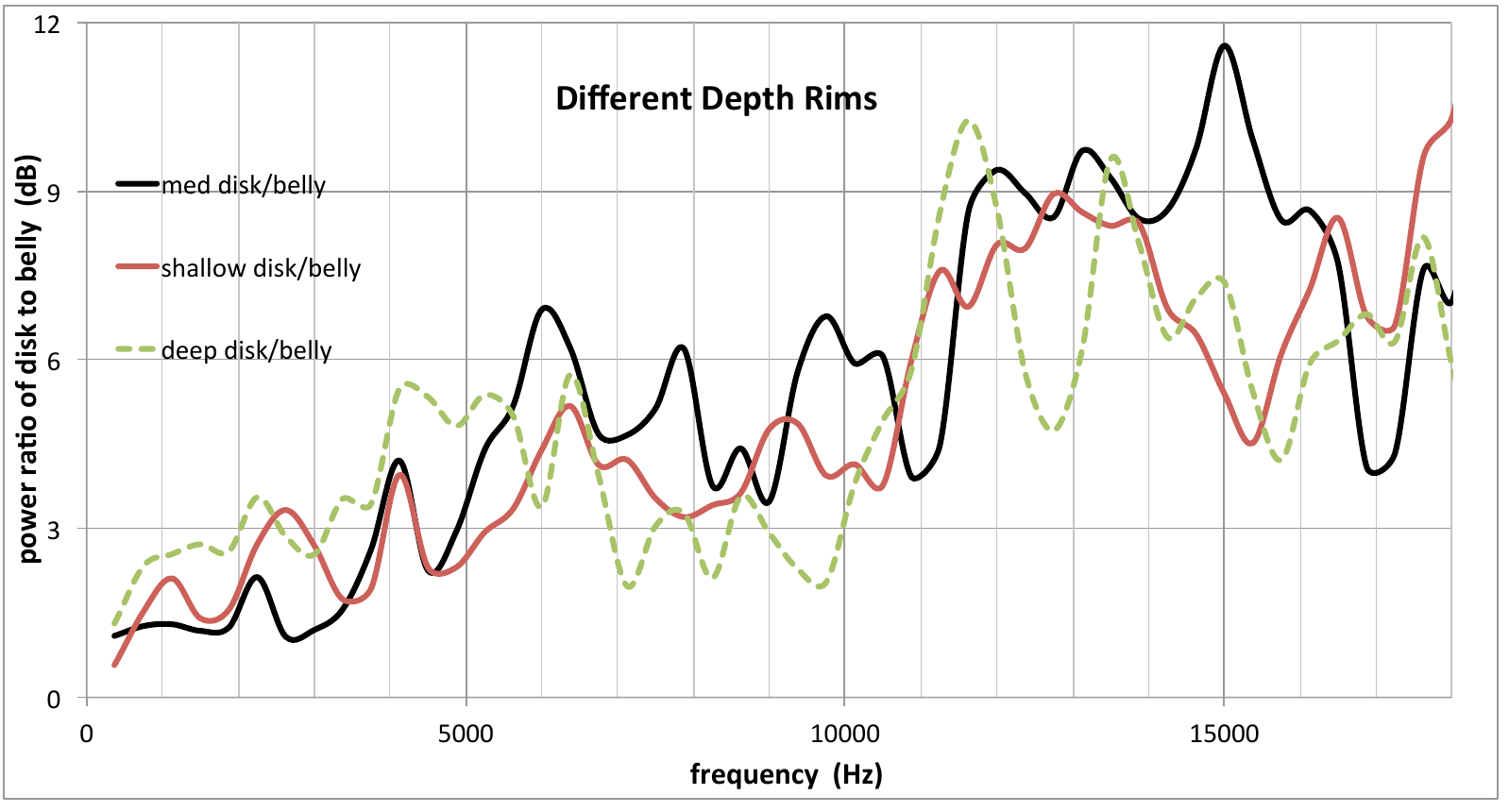}
\caption{Goodtimes with med 2 $3/4''$ rim, shallow $2''$ rim, and deep 5 $5/8''$ rim}
\end{figure}

The relevant issue is the difference between the two backs for the three banjos.    FIG.~5 shows the results for each banjo of those frequency sweeps as the ratio of the signals expressed in decibels.  Zero decibels corresponds to equal signals.

The original resonator back comparison got an extra enhancement from the microphone placement and the fact that the resonator projected towards that microphone.  (That's the geometry in which listeners [rather than players] hear the banjo.)   The disk-belly differences are not quite as dramatic, perhaps because the radiation patterns are now very similar.

The frequencies of the lowest modes with standing waves of head-to-back-air motion are 3385, 2460, and 1200 Hz --- for the shallow, medium, and deep pots, respectively.   The measurements do not show any clear evidence of relevance of those frequencies.  Hence, it appears that the frequency dependence of absorption in the belly material (having used the same belly on each banjo) is the dominant factor in the gross disk-{\it versus}-belly back comparisons.

\section{some experimental and analysis details}
Room sound and the directionality of sound projection deserve some comment.  Also, a few details about the frequency analyses were not discussed earlier, in an effort to present the basic message clearly.  Some of those are discussed here.

By ``room sound" I do not mean background noise.  Rather, room sound is the sound of the instrument as reflected off the surfaces of the room.  Some amount is present wherever you listen or record.  The delay relative to the direct sound is typically undetectable as a time interval, but the room sound adds and subtracts from the direct sound to various degrees, depending on the phase differences.  Certain frequencies will be enhanced and others reduced, depending on the location of the instrument and the microphone in the room.

Banjos are notorious for the directionality of the projection of their sound, and the directionality is strongly frequency dependent.  This is clear from casual listening as well as from simple physics considerations.  It is an obvious problem for making comparisons between instruments that project differently, as is the case with resonator and openback banjos.  And there is no unique, correct  comparison.  One possibility is the sound heard listening across a modest size room.  An extreme example is the sound averaged over an enclosing spherical surface as measured in an anechoic chamber.  The latter may make contact with some physics calculation, but no one listens to music that way.  The sound of a recording depends on microphone placement.   And mixing more than one mic (even without any filtering or equalizing) creates a continuum of possible sounds.

For the previous, lower frequency back comparison\cite{part1}, I recorded at a considerable distance, with a variety of baffles, over a circuitous route.  Here, I set the microphone rather close to reduce the ratio of room sound to direct sound and placed the mic in a place that favored the resonator over the openback because that's closer to natural listening.  The flat disk back comparisons were on a more equal footing because the banjo geometries were the same for disk and belly.

Unless there are no reflective surfaces, room sound is potentially an issue.  Close placement of the mic reduces its strength relative to the sound that comes directly from the instrument.  However, the nature of the radiated sound depends on distance.  Different frequencies spread out differently.  So a close microphone records sound which differs from what an audience hears.

Normally, the active sound processing of our brains tries to ignore room sound.  That involves calling on past experience and moving our heads and ears as we listen.  (It is also possible to focus specifically on the room sound, e.g., if you want to know where you are).  A stationary microphone (or one not connected to our body) yields a dramatically different experience, undercutting much of that normal process.  A dramatic version occurs with single, constant frequency sounds (as were used in the tweeter sweeps).  When the frequency is at a resonance of the room, the room sound becomes a standing wave with nodes and maxima at fixed locations.  For wavelengths small compared to the room dimensions, there are a great many resonances very closely spaced in frequency.  Move your head, and you hear loud and then soft.  A fixed microphone is somewhere in that standing wave field, but not necessarily at a maximum.  If the driving frequency is changed slowly, the node planes will move and cross through the stationary microphone as the frequency changes.  So the recorded volume will go up and down --- not because of going on or off a resonance but because of being near a maximum or node of the current frequency.

The moving of node planes offers a way to reduce the confusion due to room sound in the frequency sweep recordings.  With the microphone and banjo in a fixed relation, one can move them and rotate them together, repeating the same sweep, and average those recordings.  In FIG.~2, there's a comparison of a single sweep and the average of four.  And all of the disk and openback data are the average of three runs with different orientation and placement of the fixed-relation mic and banjo.  Because the phase relation between the direct sound and the room sound at the microphone is different for different locations, it is possible for the room sound to cancel out with sufficiently many different runs.  In practice, one makes a compromise, doing more than one but fewer than infinity.

In performing a numerical spectrum calculation, the finite time length of the data and the finite time available to do the calculation necessitate compromises.  That can be used to your advantage.  Audacity{\small $^{\circledR}$} makes explicit your choice of time interval to be considered for calculating the spectrum.  That choice determines the frequency resolution.  Because the current investigation studied the sound response to a sweep of driver frequency (and not to particular plucked notes), a crude resolution was appropriate. (In normal playing, the pot is called upon to respond to a great variety of frequencies.)   For the disk/belly comparisons, my choice is equivalent to about a 200 Hz resolution at all frequencies.  The resonator/belly comparison used a 100 Hz resolution for the lower half (on log scale) of the spectrum and 200 Hz for the upper.

\section{conclusion}
The characteristic resonator banjo sound may be hard to pin down in terms of mathematical description or specification.  The strength of the transduction path from string pluck to radiated sound at frequencies above 5000 Hz plays a crucial part.  Less of that energy is absorbed by a resonator than by a body behind the openback.  The resonator sound may remain in the realm famously identified by Supreme Court Justice Potter Stewart, to paraphrase ``I know it when I hear it."  The onomatopoeic  double plosive characterization that a resonator banjo can ``really crack" may be the best we can do.  But, as a diagnostic tool for design and development, very high frequency testing may prove useful. 

\appendix

\section{the Goodtimes as openbacks}

In ref.~\cite{openback} I compared the sound of the three Goodtime banjos played as open backs.  The only physics analyses there addressed the lowest two modes, i.e., the Helmholtz mode of the pot and the lowest head mode.  I developed  simple models of how these depended on the pot depth.  There were clearly other differences in the actual sounds of the banjos.  For example, my friend Rick concluded that no head stuffing was required on the deep pot.  It produced the mellow pluck sound he liked without any such modification.  The present tweeter-driven comparison allows an investigation of subtle but audible differences at high frequencies.  FIG.~6 shows the results for the three openback Goodtimes.
\begin{figure}[h!]
\includegraphics[width=6.5in]{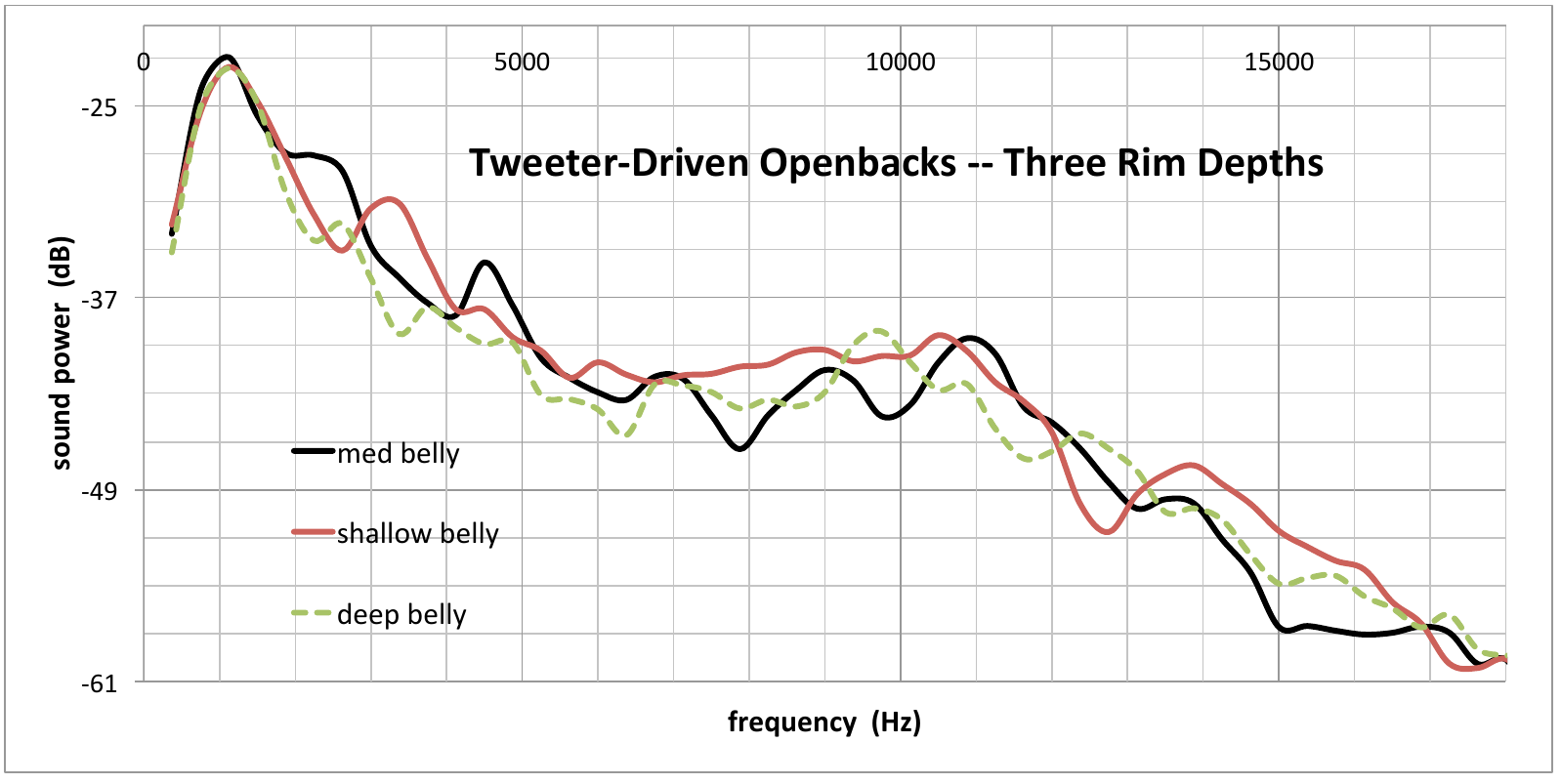}
\caption{high frequency addendum to {\it The Open Back of the Openback Banjo}\cite{openback}}
\end{figure}
One would need an understanding of what Rick was listening for in terms of frequency before being completely convinced by those measurements.  There certainly are extended regions of high frequencies where the deep pot was significantly quieter than the others.  Rim walls also dissipate sound.  For these three banjos, the wall area is proportional to their depths.  So the deep one will show more dissipation overall.  However, the ratios of the various pot depth responses are not roughly constant over the whole observed frequency range.  And increasing the pot depth does not simply shift the response to lower frequencies inversely proportionate to the pot-head echo time. That means that there is more going on than the simple story presented here.  Certainly central to that subject is the coupling of the head to the air in the pot, and that will be left for future study.

\section{sound samples: which is which}
\href{http://www.its.caltech.edu/~politzer/back-key.pdf}{Click here} or otherwise retrieve \href{http://www.its.caltech.edu/~politzer/back-key.pdf}{http://www.its.caltech.edu/$\sim$politzer/back-key.pdf} to find which sound file had a real resonator, which had synthetic resonator sound from high frequency enhancement, etc.

...or, if you only have this version, see the end of the notes and references.



\begin{thebibliography}{99}
\bibitem{tone-ring}Banjo examples of such systems that are independent to a zeroth approximation are the individual strings, the head, the rim, and, here, the pot interior.  A banjo counter example is the tone ring.  That ring is sufficiently strongly attached to the rim that it cannot vibrate independently of the top edge of the rim.  Consequently, the resonant frequencies of the ring by itself have no immediate, simple relation to the what the ring does when it is in place and the dynamics of the rim are essential.
\bibitem{nodes}In a normal mode, all the air is moving in phase (or $180^{\text{o}}$ out of phase) with a single frequency.  The pressure is oscillating likewise.  There are stationary places where the amplitude of motion is zero, and those are called nodes.  In this three-dimensional system, the nodes come in planes.  Explicit solution of sound in a cylinder as a physics exercise usually does not appear until the third year of college physics.  It can be found in any number of intermediate textbooks and other references.
\bibitem{total-internal}This is the phenomenon of total internal reflection.  At small angles to an interface, a wave in the slower medium cannot generate a wave in the faster medium.  A common example is light at a water-air interface.  It reflects completely, going back into the water, if the incident angle from the water onto the interface is sufficiently far from perpendicular.

\newpage
\end{thebibliography}
\end{document}